\documentclass[10pt,final, twocolumn]{IEEEtran}
 
\usepackage{multirow}  
\usepackage{amsmath,graphicx}
\usepackage{amssymb}
\usepackage{acronym}

\usepackage[belowskip=-10pt,aboveskip=1.8pt]{caption}
\usepackage{subcaption}
\usepackage{standalone}
\usepackage{tikz}
\usepackage{url,cite}
\usepackage{verbatim, cases}
\usepackage[bookmarks,colorlinks]{hyperref}
\usepackage{soul, xcolor, flushend}
\usepackage{soul}
\usepackage{placeins}
\usepackage{enumitem, dirtytalk}

\usepackage[ruled,linesnumbered,vlined]{algorithm2e}
\SetKwInput{KwData}{\textbf{Init}}

\definecolor{NewColor}{rgb}{0,0,0}
\definecolor{NewColor2}{rgb}{0,0,0}

\acrodef{lwa}[LWA]{leaky wave antenna}
\acrodef{los}[LOS]{line-of-sight}
\acrodef{hmm}[HMM]{hidden Markov model}
\acrodef{dl}[DL]{deep learning}
\acrodef{dnn}[DNN]{deep neural network}
\acrodef{snr}[SNR]{signal-to-noise ratio}
\acrodef{bs}[BS]{base station} 
\acrodef{cpu}[CPU]{centralized processing unit} 
\acrodef{mimo}[MIMO]{multiple-input multiple-output}
\acrodef{awgn}[AWGN]{additive white Gaussian noise} 
\acrodef{cpu}[CPU]{central processing unit} 
\acrodef{ml}[ML]{machine learning} 
\acrodef{mse}[MSE]{mean-squared error}
\acrodef{iot}[IOT]{Internet of Things}
\acrodef{rmse}[RMSE]{root mean squared error}
\acrodef{rmspe}[RMSPE]{root mean squared periodic error}
\acrodef{mmse}[MMSE]{{minimum mean-squared error}}
\acrodef{lmmse}[LMMSE]{{linear} MMSE}
\acrodef{mle}[MLE]{maximum likelihood estimation}
\acrodef{admm}[ADMM]{alternating direction method of multipliers}
\acrodef{dadmm}[D-ADMM]{distributed alternating direction method of multipliers}
\acrodef{sar}[SAR]{successive approximation register}
\acrodef{adc}[ADC]{analog-to-digital converter} 
\acrodef{dac}[DAC]{digital-to-analog converter} 
\acrodef{msb}[MSB]{most significant bit}
\acrodef{lsb}[LSB]{least significant bit}
\acrodef{sh}[S\&H]{sample-and-hold}
\acrodef{aod}[AoD]{angle of departure}
\acrodefplural{aod}[AoDs]{angles of departure}
\acrodef{ofdm}[OFDM]{orthogonal frequency-division multiplexing}
\acrodef{ofdma}[OFDMA]{orthogonal frequency-division multiple access}
\acrodef{tdma}[TDMA]{time-division multiple access}
\acrodef{ga}[GA]{genetic algorithm}
\acrodef{music}[MUSIC]{multiple signal classification}
\acrodef{doa}[DoA]{direction of arrival}
\acrodef{ris}[RIS]{reconfigurable intelligent surface}
\acrodef{rf}[RF]{radio-frequency}
\acrodef{em}[EM]{radiate electromagnetic}

\acrodef{aoa}[AoA]{Angle of Arrival}
\acrodefplural{aoa}[AoAs]{Angles of Arrival}
\acrodef{em}[EM]{electromagnetic}
\acrodef{cmos}[CMOS]{complementary metal-oxide semiconductor}
\acrodef{ttd}[TTD]{true-time-delay}

\usetikzlibrary{trees,arrows}

\setstcolor{blue}

\begin{document}

\title{Leaky Wave Antennas for Next Generation Wireless Applications in sub-THz Frequencies: Current Status and Research Challenges}

\author{
Natalie Lang, 
Atsutse K. Kludze,
Nir Shlezinger, 
Yasaman Ghasempour, 
Tirza Routtenberg, 
\\
George C. Alexandropoulos, 
and Yonina C. Eldar
\thanks{N. Lang, N. Shlezinger, and T. Routtenberg are with the School of ECE, Ben-Gurion University of the Negev, Israel (e-mail: langn@post.bgu.ac.il; \{nirshl; tirzar\}@bgu.ac.il).} 
\thanks{A. K. Kludze and Y. Ghasempour are with the Department of ECE, Princeton University, NJ, USA (e-mail: \{kludze; ghasempour\}@princeton.edu).} 
\thanks{G. C. Alexandropoulos is with the Department of Informatics and Telecommunications, National and Kapodistrian University of Athens, Greece and the Department of Electrical and Computer Engineering, University of Illinois Chicago, Chicago, IL, USA (e-mail: alexandg@di.uoa.gr).} 
\thanks{Y. C. Eldar is with the Faculty of Math and CS, Weizmann Institute of Science, Rehovot, Israel and the Department of Electrical and Computer Engineering, Northeastern University, MA, USA (e-mail: yonina@weizmann.ac.il).
}}



\maketitle

\begin{abstract}
The ever-growing demand for ultra-high data rates, massive connectivity, and joint communication-sensing capabilities in future wireless networks is driving research into sub-terahertz (sub-THz) communications. While these frequency bands offer abundant spectrum, they also pose severe propagation and hardware design challenges, motivating the search for alternative antenna solutions beyond conventional antenna arrays. Leaky-wave antennas (LWAs) have  emerged as a promising candidate for sub-THz systems due to their simple feed structure, low fabrication cost, and inherent angle-frequency coupling, which enables frequency-controlled beamsteering with simple hardware. In this article, we review the fundamentals of the LWA technology, highlight their unique properties, and showcase their potential in multi-user wideband sub-THz wireless communications. We present representative studies demonstrating that LWAs can simultaneously support high-rate multi-user communications and accurate localization using only a single antenna element. Finally, several key open challenges are outlined, spanning algorithm design, signal processing, information theory, standardization, and hardware implementation, that need to be addressed to fully harness LWAs as a cost-effective and scalable enabler of next generations of wireless systems.
\end{abstract}


\acresetall

\section{Introduction}\label{sec:into}
\IEEEPARstart{T}{he} ever-increasing demand for higher data rates, massive connectivity, and ultra-low latency in future wireless systems has driven considerable interest in sub-terahertz (sub-THz) frequency bands. These spectral regions, typically spanning $100$~GHz to $1$~THz, offer excessive chunks of bandwidth  that can facilitate meeting these growing demands~\cite{akyildiz2022terahertz}. 
However, the efficient utilization of THz spectrum poses unique difficulties, including high propagation losses, in addition to hardware challenges associated with cost-affordable circuitry operating at such increased rates~\cite{rappaport2019wireless}. The need to overcome these limitations and facilitate transmission with sufficient link budget motivates exploring antenna technologies that both enable some form of beamforming, while being cost- and power-efficient in sub-THz bands.

Conventional architectures that provide beamforming capabilities are typically based on \ac{mimo} systems in practice, which constitute a key pillar of current physical-layer technology. Yet, scaling \ac{mimo} transceivers to the THz domain can be prohibitively expensive. Fully digital MIMO architectures require a dedicated \ac{rf} chain per antenna element, leading to immense hardware and power overheads at THz frequencies~\cite{akyildiz2022terahertz}. Even hybrid \ac{mimo} architectures, which aim to strike a balance between performance and cost by combining analog and digital components~\cite{ioushua2019family}, remain highly complex and costly due to the need for large arrays of high-speed phase shifters, mixers, and amplifiers that are difficult to implement, costly, and power-hungry in the sub-THz bands. These limitations necessitate exploration of alternative antenna technologies that are intrinsically more compatible with the constraints of THz communication systems.

\Acp{lwa}, an established and relatively simple antenna technology,  have recently emerged as a promising candidate for THz applications~\cite{guerboukha2023conformal}. The suitability of \acp{lwa} for THz signaling stems from their unique radiation properties, low fabrication complexity, and angle-frequency coupling behavior. Unlike conventional \ac{mimo} arrays, \acp{lwa} steer their beams through {\em frequency variation} rather than {\em phase control}, enabling beamsteering with minimal hardware complexity, without relying on active phase shifters~\cite{comite2018analysis}. Their simple feed structure, compatibility with standard manufacturing processes, and ability to integrate with silicon-based platforms make \acp{lwa} particularly attractive for cost-sensitive and power-constrained THz systems~\cite{ghasempour2020single}. Moreover, the inherent angle-frequency coupling characteristic of \acp{lwa} can be exploited for multiplexing and spatial processing, opening new avenues for multi-user communications as well as joint communications and sensing designs in future networks.

\begin{figure}
    \centering
    \includegraphics[width=0.9\linewidth]{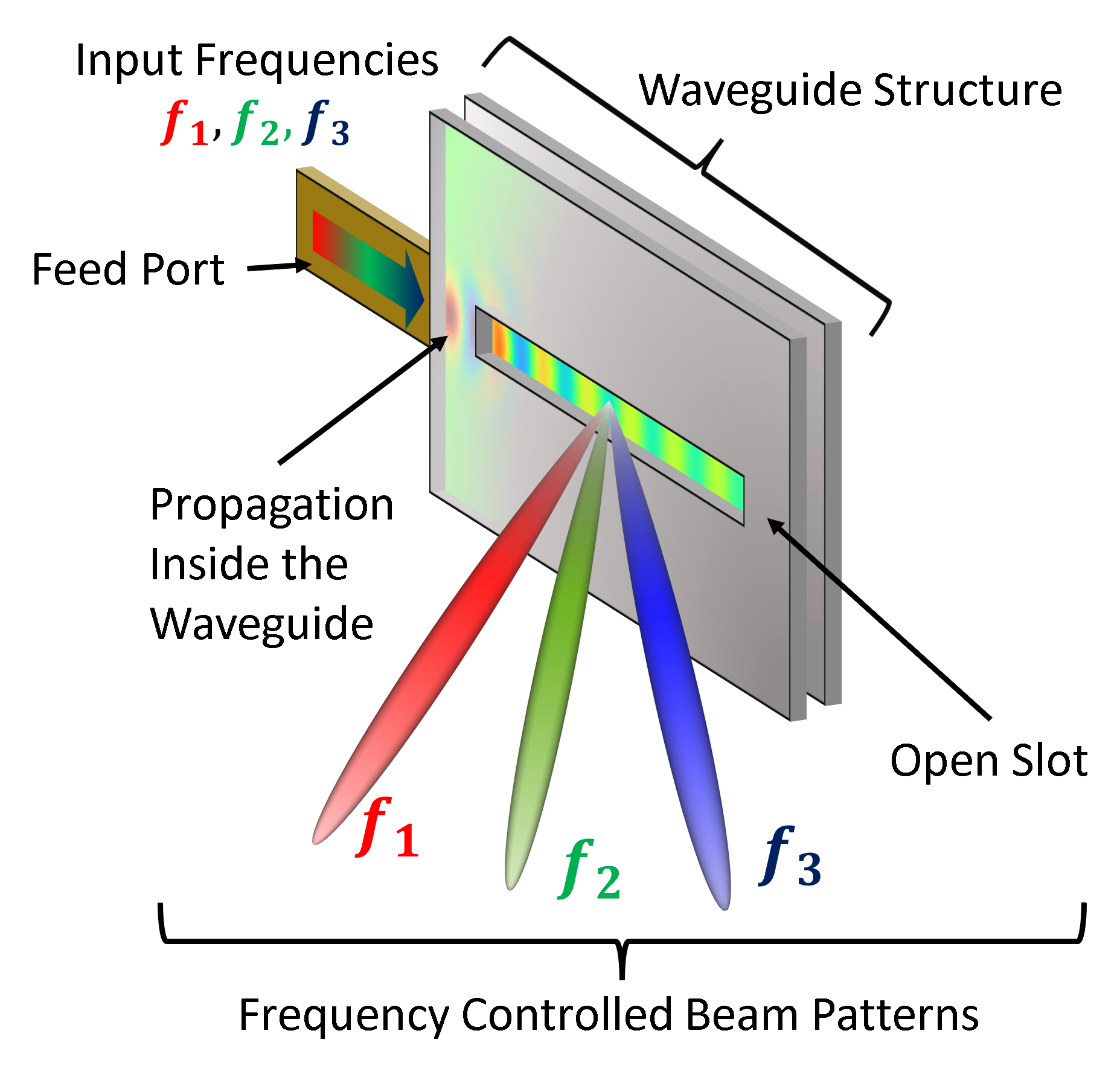}
    \caption{A parallel LWA consisting of a uniform slot and an air-wave metallic waveguide. A feed port is used to inject the waveguide with an input signal, and then, this signal is emitted out of the waveguide being decomposed into different spatial channels based on the input signal's frequencies.} 
    \label{fig:LWA High Level}
\end{figure} 

In this article, we explore \acp{lwa} as a candidate enabling technology for wideband sub-THz wireless systems that must simultaneously support multi-user communications and high-resolution sensing. We begin by reviewing the fundamental operating principles of \acp{lwa}, emphasizing their unique angle-frequency coupling and the associated modeling considerations in the sub-THz regime. Building on this foundation, we discuss how the angle-frequency coupling property may be leveraged to cost-effectively realize beamsteering and spatial multiplexing without the need for complex \ac{rf} hardware. To complement these insights, we present representative quantitative studies showcasing the potential of a single \ac{lwa} element to simultaneously achieve competitive downlink throughput performance and enable accurate uplink localization.

These results highlight that, \acp{lwa} not only reduce hardware complexity, but also provide a natural bridge between communication and sensing functionalities. Our article also identifies open challenges that need to be addressed to unlock the full potential of \acp{lwa} in the next generation of THz networks. We discuss the algorithmic and signal processing issues raised by the angle-frequency coupling, the need for new information-theoretic foundations, and the gaps in standardization and hardware development that currently hinder large-scale adoption. By consolidating existing knowledge, demonstrating concrete use cases, and outlining pressing research directions, our goal is to position \acp{lwa} within the broader context of 6G and beyond system design, and to provide the community with a roadmap for exploiting this antenna technology in future cost-effective and scalable wireless deployments.

\section{Leaky Wave Antennas}\label{sec:LWAs}
This section reviews the fundamental operating principles of \acp{lwa}. Subsection~\ref{ssec:primer} provides a concise primer on their development, while Subsections~\ref{ssec:modeling}–\ref{ssec:OperationTHz} cover modeling and sub-THz operation, respectively.

\subsection{A Primer on LWAs}\label{ssec:primer}
\acp{lwa} were first developed in the $1940$s, during early investigations of waveguides used as radiating structures.
This antenna technology belongs to the broader category of traveling-wave antennas, which are distinct from resonant antennas in that they emit \ac{em} energy gradually along a structure rather than from a single feed point. \acp{lwa}, in particular, are based on waveguides (which can be constructed in a variety of forms, including microscopic lines, dielectric slabs, and printed circuit boards \cite{liu2011modal}) where an aperture on the structure is introduced. A traveling wave propagates along the structure either in an air-filled or dielectric medium. Upon interacting with the aperture, the wave will gradually ``leak out" and radiate. 

\subsection{Key Modeling Aspects}\label{ssec:modeling}
Given that the aperture introduces a change in the waveguide's normal operation, its geometry has a direct impact on the propagation behavior of the \ac{lwa}. For example, an \ac{lwa} with a uniform slit acts as a fast-wave traveling antenna, in which the phase velocity of the \ac{em} wave inside the waveguide exceeds the speed of light in free space. In contrast, an \ac{lwa} with a periodic slit behaves as a slow-wave antenna, where the phase velocity is lower. Consequently, such changes determine the required boundary conditions on the leakage behavior from inside the waveguide to the radiating environment (in this case, free space). By solving the phase-matching requirement using Maxwell equations, it is revealed that there is a direct coupling between the angle of leakage from the open slit and the frequency of the input signal. This property, commonly referred to as the angle-frequency coupling relationship, is the defining property of \acp{lwa}.

As illustrated in Fig.~\ref{fig:LWA High Level}, the intrinsic angle-frequency coupling of an \ac{lwa} enables frequency-controlled beamsteering. In practice, this means that as the input signal frequency varies, the angle of radiation continuously shifts, causing the main beam to scan across space without requiring active phase shifters, mechanical movement, or complex feed networks. This simple yet powerful mechanism directly links the spectral content of the transmitted waveform to spatial directions, thereby turning frequency diversity into spatial diversity. 
Beyond its immediate hardware advantages, this property also opens the door to new system-level opportunities: on the communication side, different subbands can naturally be directed towards different users, facilitating parallel multi-user transmissions a la a frequency division multiple access manner, while on the sensing side, receiving energy at a given frequency may be mapped to a known radiation angle, enabling spatial inference.
Consequently, Fig.~\ref{fig:LWA High Level} not only illustrates the hardware simplicity of \acp{lwa} but also highlights their potential as a natural enabler for joint communications and sensing in sub-THz wireless networks.

\subsection{Operation at THz Bands}\label{ssec:OperationTHz}
Since their discovery, \acp{lwa} have been extensively explored in the microwave regime for applications such as frequency scanning radar and tracking systems. More recently, they have gained attention as an alternative beamsteering strategy to conventional phased arrays. Note that the significant loss of semiconductor switches above $100$ GHz prohibits large-scale array demonstration which is actually needed to enable links under severe path loss in this regime~\cite{fu2020terahertz}. In fact, the development of on-chip antenna arrays above $100$ GHz remains an open challenge, with only a handful of successful demonstrations reported in recent literature~\cite{jalili20190}. 
Furthermore, such phase array architectures are unable to fully support the entire bandwidth of the newly available spectrum. 

\acp{lwa} instead hold many advantages for future implementation of THz communication systems. First, the large available bandwidth in the THz regime is aligned with the potential of frequency-controlled beam scanning. Second, \acp{lwa} are easy to fabricate, compact in design, and of low cost. Recent implementations of THz \acp{lwa} have been demonstrated for single-shot localization~\cite{ghasempour2020single}, low-power wideband data backscattering~\cite{kludze2024frequency}, frequency division multiplexing~\cite{karl2015frequency}, and physical-layer authentication~\cite{kludze2022towards}. 

\begin{figure}
    \centering
    \includegraphics[width=1\linewidth]{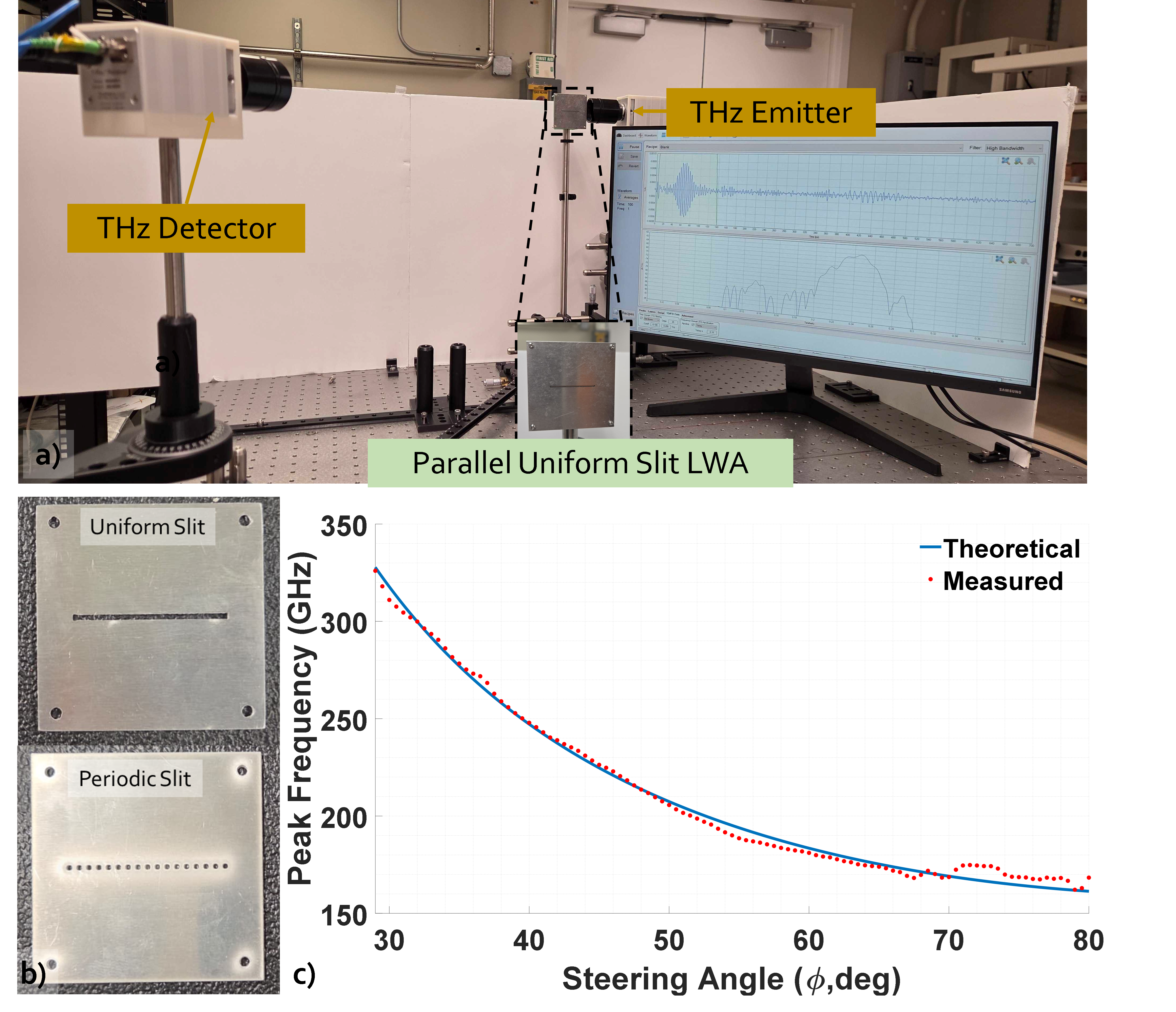}
    \caption{Implementation and prototyping of sub-THz \acp{lwa}: 
    a) A parallel plate \ac{lwa} in an experimental lab setting from the Smart Wireless Agile Networks Lab at Princeton. The \ac{lwa} is fed by a THz emitter that injects a wideband signal into the parallel waveguide, from which it leaks out of the aperture to create a frequency-dependent beam. A THz detector is placed to measure the spatial-dependent waveform. b) Prototype \ac{lwa} architectures with various aperture designs for different applications. c) Comparison between the theoretical and experimental angle-frequency coupling relationship of a uniform slit \ac{lwa}.}
    \label{fig:LWA Experiment}
\end{figure}

Fig.~\ref{fig:LWA Experiment} provides an illustration of how \acp{lwa} at THz frequencies are realized and tested in practice, bridging the gap between theoretical models and physical implementations. Part a) depicts a parallel-plate \ac{lwa} in an experimental laboratory setup, where a wideband THz emitter injects signals into the waveguide, which then leak through the aperture to produce frequency-dependent beams that can be captured by a THz detector at different spatial locations. This setup highlights how the fundamental angle-frequency coupling manifests in practice and validates the theoretical predictions of \ac{lwa} behavior. Parts (b) and (c) extend this view by demonstrating a variety of aperture designs and comparing experimental results with analytical models. The diversity of architectures in part (b) emphasizes the flexibility of \acp{lwa}: depending on the choice of aperture geometry and material platform, one can tailor the leakage characteristics to specific applications such as high-resolution localization, backscattering, or multiplexing. Finally, part (c) confirms the close agreement between theory and measurements of the angle–frequency coupling for a uniform slit \ac{lwa}, underscoring both the robustness of the underlying physics and the practicality of implementing \acp{lwa} at sub-THz frequencies. Altogether, Fig.~\ref{fig:LWA Experiment} exemplifies the experimental feasibility of \acp{lwa} for sub-THz wireless applications.

\section{LWA-Aided Wireless Systems}\label{sec:LWAs for WCOM}
This section explores the potential of an \ac{lwa} to facilitate multi-user communication networks. We commence with a qualitative modeling of an \ac{lwa} for wireless systems in Subsection~\ref{ssec:multi-user with LWAs}, focusing on its unique dispersion property, allowing for cost-effectively supporting simultaneous transmissions and sensing in the sub-THz regime. We then provide a quantitative study for representative settings of downlink beamforming and uplink localization in Subsection~\ref{ssec:experiments}.

\subsection{Multi-User Communications}\label{ssec:multi-user with LWAs}
As discussed in Section~\ref{sec:LWAs}, \acp{lwa} exhibit an intrinsic frequency-dependent steering effect: as the frequency varies, the main beam of the antenna shifts accordingly. This results in the radiated energy being inherently spread over a range of directions rather than concentrated in a narrow beam. In conventional point-to-point wireless communications, this may be seen as a drawback, as much of the signal power is \say{wasted} outside the intended path.
Yet, in scenarios where users are spatially distributed, this natural dispersion aligns well with the geometry of the environment, e.g, in multi-user systems. There, the behavior of an \ac{lwa} is not only well suited, but also features low-cost and simple hardware that stands in contrast to prevailing alternatives, particularly when scaled for sub-THz operation.

\begin{figure*}
		\centering
        \begin{subfigure}{0.42\textwidth}
        \centering
            \includegraphics[width=0.9\linewidth]{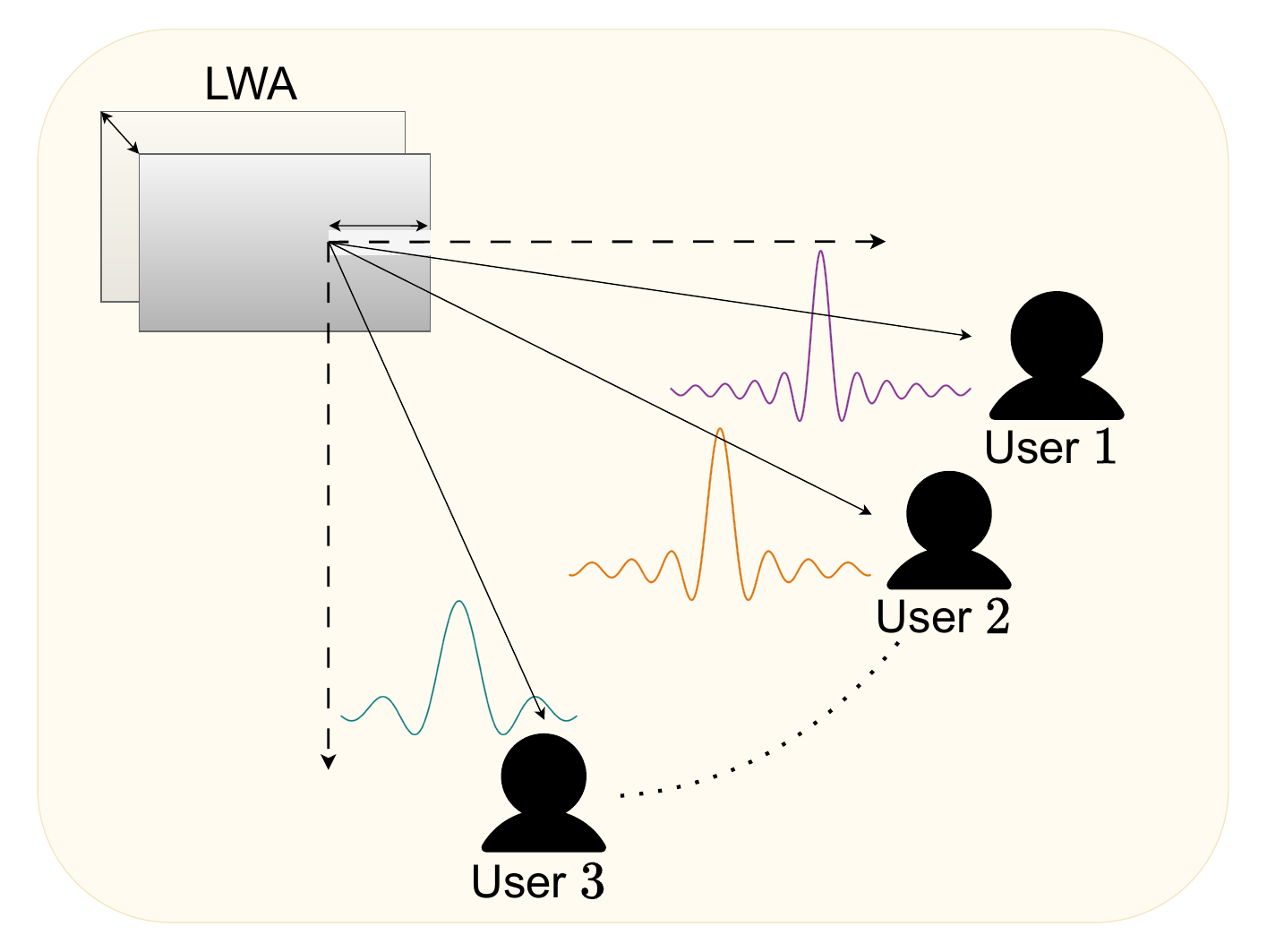}
         \end{subfigure}
        \begin{subfigure}{0.57\textwidth}
        \centering
            \includegraphics[width=0.9\linewidth]{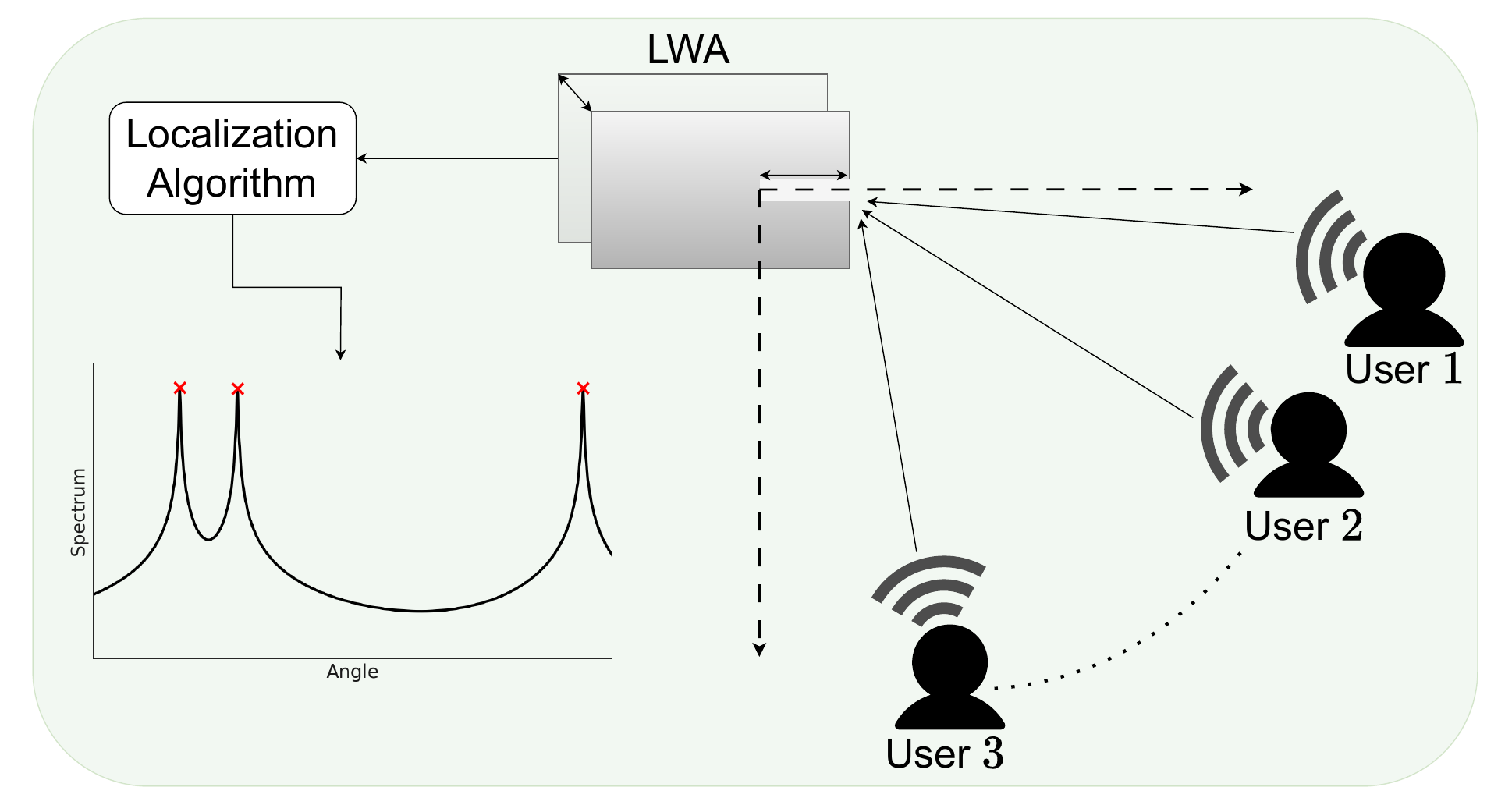}
        \end{subfigure}
    \caption{A single \ac{lwa} supporting multiple users for downlink beamforming (left) and uplink localization (right).} 
    \label{fig:lwa_usages}
\end{figure*}

More specifically, the fundamental physics of an \ac{lwa} allows accurate modeling of its beamforming characteristics, mapping each radiated frequency component to a distinct angle and beamwidth. This property can be leveraged to provide parallel signaling to multiple users using a single antenna element, as opposed to traditional \ac{mimo} architectures. The passive distribution of the frequency band across space in an \ac{lwa} can be further optimized by jointly adjusting the users’ power and spectrum allocation \cite{lang2025widebandthzmultiuserdownlink}, and, in some implementations, the antenna’s physical parameters (plate separation and slit length)~\cite{javanbakht2021review}. Beyond communication, the angle-frequency coupling of an \ac{lwa} extends to sensing applications, such as user localization, since receiving a signal at a given frequency implies a known radiation angle, enabling spatial inference. These usages, quantitatively exemplified in the sequel, are illustrated in Fig.~\ref{fig:lwa_usages}, which shows a single \ac{lwa} serving multiple spatially distributed users for downlink beamforming and uplink localization.

\subsection{Quantitative Study}\label{ssec:experiments}
We next quantitatively showcase the ability  of a single \ac{lwa} to support sub-THz multi-user wireless communication and spatial inference. To that end, we have simulated a channel spanning $[0.2, 0.8]$~THz, a range known to be feasible for THz \acp{lwa}~\cite{ghasempour2020leakytrack}. We begin by illustrating the \ac{lwa}'s angle-frequency response through its beampatterns, then assess the system throughput as a function of \ac{snr}, and finally demonstrate localization capability via \ac{music}-based angle estimation.

\subsubsection{Beampatterns}
\begin{figure}
    \centering
    \includegraphics[width=0.9\linewidth]{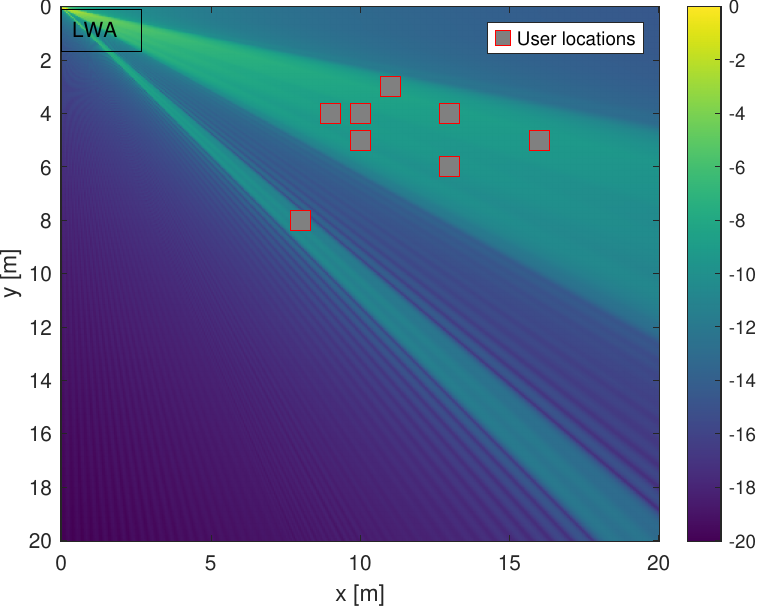}
    \caption{Log-scaled normalized energy of directed beams towards $8$ users using  an \ac{lwa}-based \ac{bs}.} 
    \label{fig:beams_OFDM}
\end{figure}

To evaluate the performance of an \ac{lwa}-aided \ac{bs} in generating directional beams, we consider an \ac{lwa} excited with an \ac{ofdm} waveform, in a medium-scale scenario with $8$ randomly positioned users, where we tune the spectral power allocation along with the \ac{lwa} configuration to maximize the achievable rates. 
\textcolor{black}{In this setup, different users are served by radiating distinct frequency components toward different spatial directions corresponding to each user.}
Fig.~\ref{fig:beams_OFDM} illustrates the resulting log-scaled normalized energy that radiates towards each position over the entire spectrum.
It can be observed in  Fig.~\ref{fig:beams_OFDM} that a single \ac{lwa} can steer distinct beams towards multiple users, utilizing the inherent angular dispersion of frequency components, where higher frequencies are naturally directed toward smaller angles with narrower beams. This ability to simultaneously serve multiple users with a single-element antenna underscores the potential of \acp{lwa} as a cost-effective solution for sub-THz multi-user wireless systems.

\subsubsection{Sum-rate vs. SNR}
\begin{figure}
    \centering
     \includegraphics[width=0.9\linewidth]{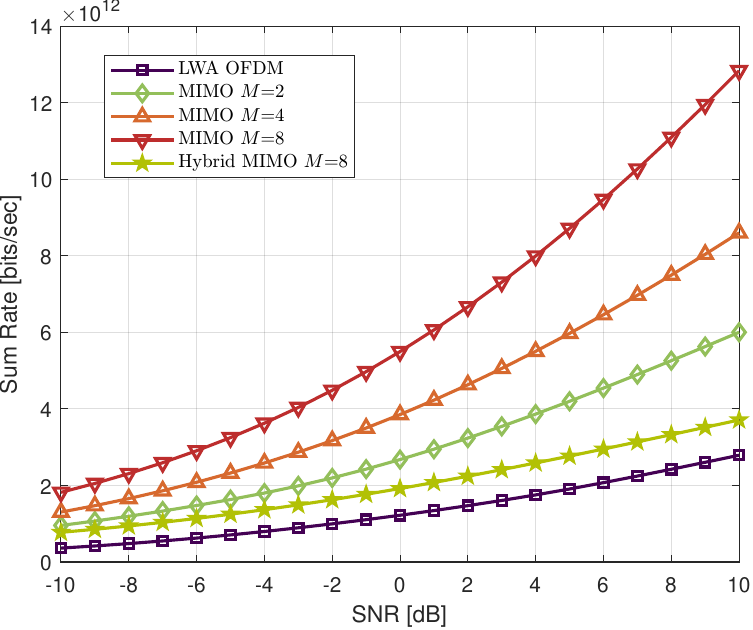}
    \caption{Sum-rate performance versus \acs{snr} for $32$ users, comparing \ac{lwa} with $M$-antenna digital/hybrid \ac{bs}.} 
    \label{fig:LWA_vs_MIMO}
\end{figure}

Building on these observations, we next assess system-level throughput performance in Fig.~\ref{fig:LWA_vs_MIMO}. There, we compare the achievable sum-rate performance versus \ac{snr} of an \ac{lwa} and conventional \ac{mimo} architectures in a large-scale scenario with $32$ users. The baselines include $(i)$ a fully digital \ac{mimo} configuration with $M$ antennas; and $(ii)$ a hybrid analog/digital array employing a single RF chain with phase shifters optimized as in~\cite{ioushua2019family}. Both \ac{mimo} architectures are notably more costly, involving multiple elements and active RF chains at THz bands.
\textcolor{black}{There, multiple users are served simultaneously through spatial multiplexing enabled by precoding.}
For a fair comparison, all systems are evaluated under identical conditions, i.e., the same user locations and \ac{snr} levels. 
As shown in Fig.~\ref{fig:LWA_vs_MIMO}, the \ac{lwa} alternative, which operates with a single element with inherent angle-frequency dispersion, achieves competitive performance relative to the more complex \ac{mimo} setups, further demonstrating its practicality in scalable deployments of wideband wireless networks.

\subsubsection{Localization}
\begin{figure}
    \centering
     \includegraphics[width=0.9\linewidth]{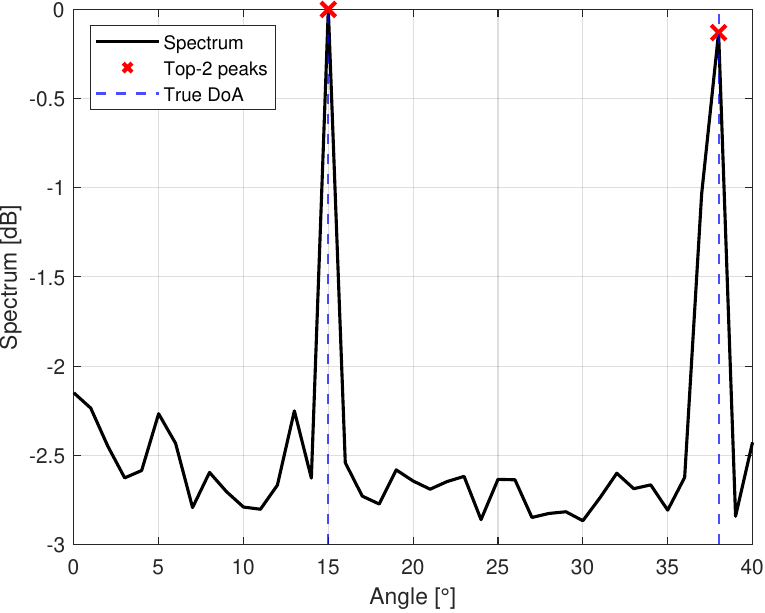}
    \caption{LWA MUSIC Spectrum ($\text{SNR}=0$ [dB]).} 
    \label{fig:MUSIC_spectrum}
\end{figure}

To assess how the spectral-spatial coupling of \acp{lwa} can be leveraged for sensing purposes, we simulate a wideband uplink scenario where two users transmit pilot symbols from distinct angular positions. Although the receiver here has a single antenna element, it can construct the \ac{music} spectrum, shown in Fig.~\ref{fig:MUSIC_spectrum}, from a single-snapshot wideband measurement. As opposed to the conventional formulation of \ac{music} in array signal processing, which constructs a spatial spectrum from multiple elements using the array response (steering matrix), the spectral-spatial coupling induced by \acp{lwa} allows forming a \ac{music} spectrum from multiple frequency measurements using the \ac{lwa} steering model.
The spectrum exhibits distinct peaks at the true \acp{doa}, even in the presence of noise and without relying on a conventional antenna array.
This result confirms that the inherent angle-frequency coupling in \acp{lwa} enables subspace-based algorithms such as \ac{music} to resolve multiple user directions using only a single spatial element. This capability allows accurate angle estimation without the complexity of phased arrays or fully digital beamforming architectures, positioning \acp{lwa} as an efficient and compact solution for user localization in sub-THz systems.

\subsubsection{Summary}  
The results of our quantitative evaluation highlight the unique advantages that \acp{lwa} bring to wideband multi-user wireless systems. First, the beampattern analysis demonstrates that a single \ac{lwa} inherently radiates frequency components into distinct angular directions, enabling the formation of multiple beams simultaneously without resorting to active phase shifters or large antenna arrays. This natural angle-frequency dispersion, when coupled with optimized power and spectrum allocation, allows one to effectively serve multiple users with competitive throughput performance. Indeed, the sum-rate analysis indicates that an \ac{lwa}-based \ac{bs} operating with a single radiating element can approach the performance of considerably more complex MIMO configurations, underscoring the potential of \acp{lwa} as a cost- and energy-efficient alternative for sub-THz networks.

Equally important are the sensing capabilities revealed by the localization study. By exploiting the deterministic angular mapping of \acp{lwa}, subspace-based algorithms such as MUSIC can reliably infer the directions of arrival of multiple users from wideband snapshots, even when relying on only a single antenna element. This dual capability---simultaneously supporting high-throughput communications and accurate spatial inference---positions \acp{lwa} as a uniquely powerful technology for future joint communication and sensing systems. Overall, the presented results indicate that \acp{lwa} are not only a pragmatic solution to the hardware challenges of sub-THz beamforming but also a natural enabler of integrated multi-user communication and sensing functionalities for next-generation wireless networks.

\section{Research Challenges and Opportunities}\label{sec:challanges} 
Despite the promising capabilities of \acp{lwa} in enabling low-cost, energy-efficient, and beamsteerable transceivers for sub-THz wireless systems, several key challenges must be addressed before they can be incorporated in wireless communication networks. These challenges span multiple layers of the system design, including network-level orchestration, signal processing, theoretical modeling, standardization, and hardware implementation. Bridging the gap between \ac{lwa}-based physical-layer functionalities and their integration into the broader sixth-generation (6G) wireless stack requires addressing these open problems through dedicated research. In the following, we outline key directions where future work is needed to fully harness the potential of \acp{lwa}.

\subsection{Algorithms for Communications}
The integration of \acp{lwa}  introduces novel coordination and control challenges. First, the inherent angle-frequency coupling characteristic of \acp{lwa} implies that changes in user location can lead to significant variations in the spectral profile. This necessitates the development of robust synchronization mechanisms tailored to the \ac{lwa} propagation model. Second, mechanisms are needed to facilitate coexistence and dynamic collaboration between conventional \ac{mimo} transceivers and \ac{lwa}-based nodes. Such mechanisms should include intelligent selection policies that determine the most suitable antenna architecture and frequency band based on environmental conditions and user location. Finally, combining \acp{ris} with \acp{lwa} holds promise for extending the angular coverage of single \ac{lwa} elements. However, efficient integration strategies that jointly optimize \ac{ris} configurations and \ac{lwa} parameters remain largely unexplored.

\subsection{Signal Processing Design}
The design of signal processing algorithms that fully leverage the characteristics of LWAs is still in its infancy. New waveform designs are needed that account for and exploit the radiation profiles and frequency-angular coupling induced by LWA structures. Moreover, the unique channel geometry necessitates the development of specialized beam and frequency allocation algorithms that consider the spectral selectivity of user directions. LWA-specific channel estimation and tracking schemes must also be developed, as conventional techniques designed for flat spatial responses may fail to operate reliably under the highly frequency-selective angular behavior of LWAs.

\subsection{Information Theoretic Analysis}
Information-theoretic foundations are needed for understanding the fundamental performance limits of \ac{lwa}-enabled communication systems. Characterizing the capacity regions of channels induced by \acp{lwa}, particularly in multi-user and frequency-selective settings, remains an open problem. In addition, since \acp{lwa} can support joint communications and localization, fundamental bounds on achievable localization accuracy under \ac{lwa} constraints must be derived, especially when considering practical impairments such as limited beamwidth, bandwidth constraints, and hardware imperfections. 
These questions can also be approached from an estimation-theoretic perspective. The deterministic link between frequency and angle implies that Cramér–Rao bounds (CRBs) can describe localization accuracy. Exploring how such bounds relate to achievable rates and whether tighter CRBs correspond to greater spatial degrees of freedom could reveal new joint information–estimation limits for integrated sesning and communications with LWAs.

\subsection{Hardware Design}
Practical \ac{lwa} adoption requires  the development of reconfigurable and rapidly tunable hardware platforms. Implementing \acp{lwa} with real-time frequency agility, wideband operation, and compact form factors remains an active area of research in antenna and materials science. Furthermore, developing scalable fabrication processes for producing high-performance \acp{lwa} that can be integrated with existing silicon-based systems is critical for their cost-effective deployment in commercial wireless networks.

\subsection{Standardization}
To support widespread deployment of \ac{lwa}-based systems in future THz wireless networks, standardized frameworks are required that account for the distinct operational features of \acp{lwa}. These include the modeling of angle-frequency coupling in physical-layer abstractions, the specification of control signaling for beam selection and frequency tuning, and the definition of interoperable interfaces between \acp{lwa}-equipped nodes and the broader network infrastructure. Standardization efforts must also consider how to accommodate \acp{lwa} within the native AI and open radio access network (RAN) paradigms being developed for 6G, enabling flexible, data-driven optimization of \ac{lwa}-based links.

\section{Conclusions}
In this article, we reviewed \acp{lwa} as a promising and cost-effective antenna technology for future sub-THz wireless networks. We began by outlining the fundamental operating principles of \acp{lwa} and highlighting their distinctive angle-frequency coupling property, which enables simple frequency-controlled beamsteering. Building on these foundations, we examined how \acp{lwa} can be harnessed to support multi-user communications as well as joint communications and sensing functionalities, presenting quantitative studies that demonstrate their ability to achieve competitive sum-rate performance and accurate user localization using only a single antenna element. We further discussed the practical challenges spanning algorithm design, signal processing, theoretical modeling, hardware designs, and standardization, which need to be addressed to fully realize the potential of \acp{lwa}. Taken together, our article positions \acp{lwa} as a compelling candidate technology for enabling scalable, energy-efficient, and integrated sensing and communication solutions in next generations of sub-THz wireless systems.

\bibliographystyle{IEEEtran}
\bibliography{IEEEabrv,bib}

\end{document}